\documentclass[12pt]{iopart}
\usepackage{epsf}
\usepackage{graphicx}
\def\simlt{\lower.5ex\hbox{$\; \buildrel < \over \sim \;$}}

\def\be{\begin{eqnarray}}
\def\ee{\end{eqnarray}}

\def\ab{{\alpha-\beta}}

\begin{document}

\title[The $\alpha-\beta$ relation]
{The velocity anisotropy - density slope relation}

\author{Steen H. Hansen \& Joachim Stadel}
\address{University of Zurich, Winterthurerstrasse 190, 8057
Zurich, Switzerland} 

\begin{abstract}
One can solve the Jeans equation analytically for equilibrated dark
matter structures, once given two pieces of input from numerical
simulations.  These inputs are 1) a connection between phase-space
density and radius, and 2) a connection between velocity anisotropy
and density slope, the $\alpha-\beta$ relation. The first (phase-space
density v.s. radius) has already been analysed through several different
simulations, however the second ($\ab$ relation) has not been
quantified yet. We perform a large set of numerical experiments in
order to quantify the slope and zero-point of the $\ab$ relation.  We
find strong indication that the relation is indeed an attractor.  When
combined with the assumption of phase-space being a power-law in
radius, this allows us to conclude that equilibrated dark matter
structures indeed have zero central velocity anisotropy $\beta_0
\approx 0$, central density slope of $\alpha_0 \approx -0.8$, and
outer anisotropy of $\beta_\infty \approx 0.5$.
\end{abstract}




\maketitle

\section{Introduction\label{sec:intro}}

We have seen remarkable progress in the understanding of pure dark
matter structures over the last few years. This was triggered by
numerical simulations which have observed general trends in the
behaviour of the radial density profile of equilibrated dark matter
structures from cosmological simulations, which roughly follow an NFW
profile~\cite{nfw96,moore,Fukushige1997,Moore1998,Moore1999pro,Ghigna1998},
see also \cite{Fukushige2004,Tasitsiomi2004,Navarro2004,Reed2004,diemand04,power,Diemand:2005wv} for references.
General trends in the radial dependence of the velocity anisotropy have
also been suggested~\cite{cole,carlberg}.  Recently, more complex 
relations have been identified, holding even for systems that
do not follow the simplest radial power-law behaviour in density. These
relations are first that the phase-space density, $\rho/\sigma^3$, is a
power-law in radius~\cite{taylor}, and second that there is a linear
relationship between the density slope and the
anisotropy~\cite{hm}.
A connection between the shape of the velocity distribution
function and the density slope has also been suggested~\cite{students,HMZS}.

Using the Jeans equation together with the fact that phase-space
density is a power-law in radius allows one to find the density slope
in the central region numerically~\cite{taylor}, and even analytically
for power-law densities~\cite{jeanspaper}.  These results were
generalized in \cite{austin}.  Recently a more refined study \cite{dm}
using both the phase-space density being a power-law in radius and
also the linear relationship between density slope and anisotropy,
showed that one can solve the Jeans equations analytically and extract
the radial dependence of density, anisotropy, mass etc.

It therefore appears that we only have to understand the two numerical
relationships described above in order to fully quantify pure dark matter 
halo structures.  However, no theoretical explanation for the origin of 
these relations is known to us. 

The relationship between phase-space density and radius has been
considered several
times~\cite{taylor,jeanspaper,austin,dm,ascasibar,rasia,barnes} and
seems to be well established, however, the other crucial ingredient in
the analysis, namely the linear relationship between density slope and
anisotropy has only been investigated qualitatively~\cite{hm}.  In
this paper we perform a large set of simulations in order to quantify
this relationship. We show that with present day simulations there
does indeed appear to be a linear relation between density slope and
anisotropy. When combined with the assumption of phase-space being a
power-law in radius this implies zero anisotropy near the center with
density slope of approximately $-0.8$, and that the outer anisotropy
is radial and close to $+0.5$.

\section{Head-on collisions}
\label{sec:head}
The first controlled simulation is the head-on collision between two
initially isotropic NFW structures.  For the construction of the
initial structures, we use the Eddington inversion method as described 
in~\cite{stelios}.  We create an initial NFW structure with zero
anisotropy containing 1 million particles.  The parameters are chosen
to correspond to a total mass of $10^{12}$ solar masses (concentration
of 10), with half of the particles within 160 kpc. The structure is
in equilibrium in the sense that when evolving such a structure in
isolation its global properties remain unchanged.

We now place two such structures very far apart with 2000 kpc between the
centres, which is well beyond the virial radius. Using a softening of
0.2 kpc, we now let these two structures collide head-on with an initial
relative velocity of 100 km/sec towards each.  After several crossings
the resulting blob relaxes into a prolate structure.  We run all
simulations until there is no further time variation in the radial
dependence of the anisotropy and density.  We check that there is no
(local) rotation.  We run this simulation for 150 Gyrs (corresponding 
rougly to 10 Hubble times), which means
that a very large part of the resulting structure is fully
equilibrated.

All simulations were
carried out using PKDGRAV, a multi-stepping, parallel
code~\cite{stadel01}, which uses spline kernel softening and
multi-stepping based on the local acceleration of particles.
Force accuracy is set by an opening angle of $\Theta = 0.7$ 
which combined with the use of a 4-th order multipole expansion
results in typical RMS force errors of better than 0.1\%.
The simulations were performed on the 
zBox and
zBox2 at the University of Zurich
\footnote{\tt http://www-theorie.physik.unizh.ch/\~{}stadel/zBox}.

\begin{figure}[htb]
\begin{center}
\epsfxsize=13cm
\epsfysize=10cm
\includegraphics[width=0.8\textwidth]{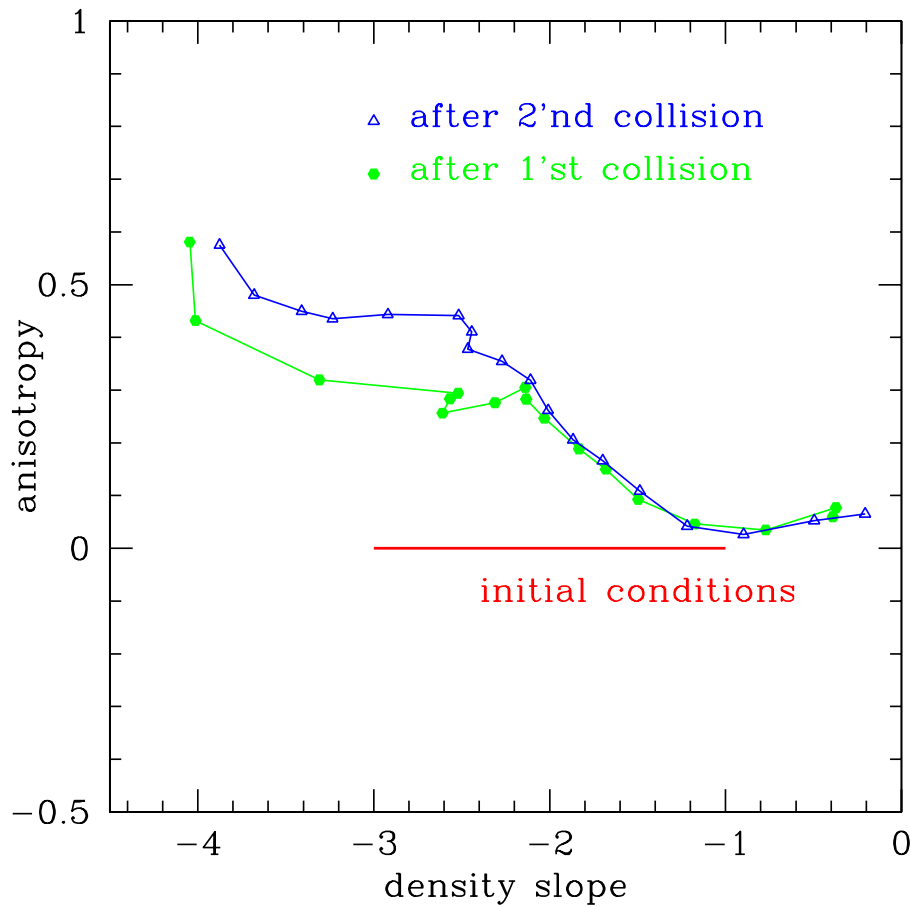}
\caption{Head-on collision between two initially isotropic NFW 
structures. The red (straight) line shows the initial condition for each of the
colliding structures. The green filled pentagons show the result after the 
first collision, where the structure contains 2 million particles. 
The blue open triangles show the result after a second collision 
(4 million particles). The green and blue symbols land very near each other 
in the region inside of a density slope of about $-2.2$}
\label{fig:headon}
\end{center}
\end{figure}

We now extract all the relevant parameters in radial bins, 
logarithmically distributed from the softening length to beyond
the region which is fully equilibrated. The resulting density
profile is very similar to an NFW profile. We  calculate the radial
derivative of the density (the density slope)
\begin{equation}
\alpha \equiv \frac{d {\rm ln} \rho}{d {\rm ln} r} \, ,
\end{equation}
and the velocity anisotropy
\begin{equation}
\beta \equiv 1 - \frac{\sigma^2_t}{\sigma^2_r} \, ,
\end{equation}
in each radial bin, where the $\sigma^2_{\rm t,r}$ are the one
dimensional velocity dispersions in the tangential and radial
directions respectively.  This is shown in figure~\ref{fig:headon} as
green filled pentagons. We also present the initial conditions as a
red (straight) line. It is clear from this figure that some connection
between $\alpha$ and $\beta$ exists, in the sense that a density slope
of about -1 has small anisotropy, and density slope of about -3 has
large radial anisotropy. In these figures we are including points
which are inside the resolved region and also points which are outside
the fully equilibrated region. We will discuss this issue fully 
in section~\ref{sec:trust}.

The first important question to address is if the $\ab$ relation
found after this collision represents an optimal configuration
of the dark matter system, or if subsequent collisions might lead to
a different configuration. We therefore take the resulting structure
(now containing 2 million particles, in a prolate shape) and 
collide 2 copies together, again separated by 2000 kpc, and with
initial relative velocity of 100 km/sec. We now use softening of
1 kpc to speed up the simulation.
The resulting structure is even more prolate when observed in density 
contours, and we evolve this for another 150 Gyr.
It should be noted that already after a few crossing times the structures
are sitting near the $\alpha-\beta$ relation, and the fact that we
run the simulations for much longer times has only a small effect
in the outer region, where the particle density is much smaller.

We again extract in radial bins, and the $\ab$ is shown as blue open
triangles in figure~\ref{fig:headon}. We see very clearly, that for
density slopes more shallow than about $-2.2$, there is virtually no
change in the $\ab$ relation. We conclude that in the central region,
where a significant perturbation occurred during
the collision, the $\ab$ relation is unchanged, and hence the
structure has indeed reached an optimal state. In the outer region
there has been a small change in the connection between $\alpha$ and
$\beta$, in the sense that the first collision brought the outer
region away from the initial conditions and the second collision
brought the system even further away.

\subsection{Dependence on shape?}
The resulting structures described above are prolate when observed in
density contours. One may fear that the resulting $\ab$ relation will
depend strongly on the axis-ratios of these structures. Naturally, if
the relevant axis-ratios  are instead the ones extracted when considering
contours in potential energy, then we should expect only an effect in
the very central region, since the potential energy contours are close
to spherical in the outer region.

\begin{figure}[htb]
\begin{center}
\epsfxsize=13cm
\epsfysize=10cm
\includegraphics[width=0.8\textwidth]{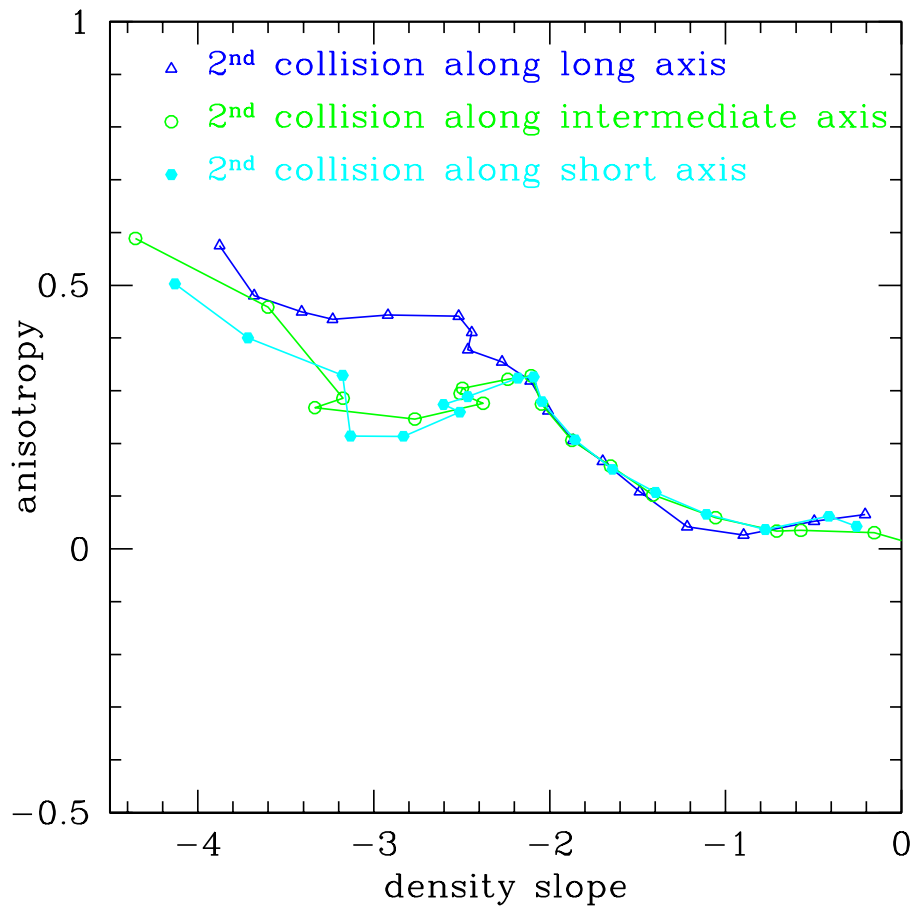}
\caption{Repeated head-on collision between two initially isotropic
NFW structures, to test the effect of shape. The different colours
refer to the axis along which the second collision was made. In the
inner region there is virtually no difference in the resulting
$\ab$ relations.}
\label{fig:shape}
\end{center}
\end{figure}

In order to test this question, we extracted the resulting prolate
structure after the first collision described above. We can now decide
to collide this structure along different axes, either along the long,
intermediate or short axes.  We perform 2 test collisions, one is a
second collision along the intermediate axis, and the other is a
second collision along the short axis (there is very little difference
between the short and intermediate axis after the first
collision). These structures end up being triaxial, almost oblate. 
The softening of these last collisions was chosen to be 2 kpc
in order to speed up the simulation.
The collision along the long axis was described above, resulting in a very
prolate structure.  The results are shown in figure~\ref{fig:shape}.

It is clear from figure~\ref{fig:shape} that in the inner region
(inside density slope of about $-2.2$) there is very little difference
between the 3 different structures. These 3 resulting structures have
rather different shape when seen in density contours, but are very
similar when seen in potential energy contours.  We conclude from
this that whereas the definition of density slope does depend
slightly on the shape of the density contours, then the $\ab$ relation
is almost independent of the shape~\cite{parisconf}.

We note that the structures should be more perturbed when collided
along the long axis, and we do indeed observe that the changes in the
$\ab$ relation in the outer region (beyond density slope of $-2.2$)
are larger when collided along the long axis as compared to collisions
along the short axis.

\subsection{Dependence on initial conditions?}

To address the question of how strongly the $\ab$ relation from the
head-on collisions described above depend on the initial conditions,
we now consider 2 simulations each with two steps.  1a) First we create
a spherical isotropic NFW structure with 1 million particles as
described above.  We take each individual particle in this structure
and put its total velocity along the radial direction, maintaining the
sign with respect to inwards or outwards from the halo centre.  We
keep the energy of each particle fixed. This structure is thus
strongly radially anisotropic.  1b) Now we take two such radially
anisotropic structures and place their centres 2000 kpc apart (well
outside the virial radius) with relative velocity of 100 km/sec
towards each other.  We use softening of 1 kpc for these tests.
Before the centres of the two structures collide the radial motion of
the particles behaves almost like a radial infall simulation, and
strong scattering of the particles is observed.  The 2 individual
structures pick random orientations in space as they become
triaxial~\cite{merritt85}; their orientations are completely erased
after the two structures collide head-on. We let the structure
equilibrate. We take two copies of the resulting
structure and collide these head-on, again with 100 km/sec. The final
equilibrated region contains approximately 2.5 million particles (of
the total of 4 million particles).

\begin{figure}[htb]
\begin{center}
\epsfxsize=13cm
\epsfysize=10cm
\includegraphics[width=0.8\textwidth]{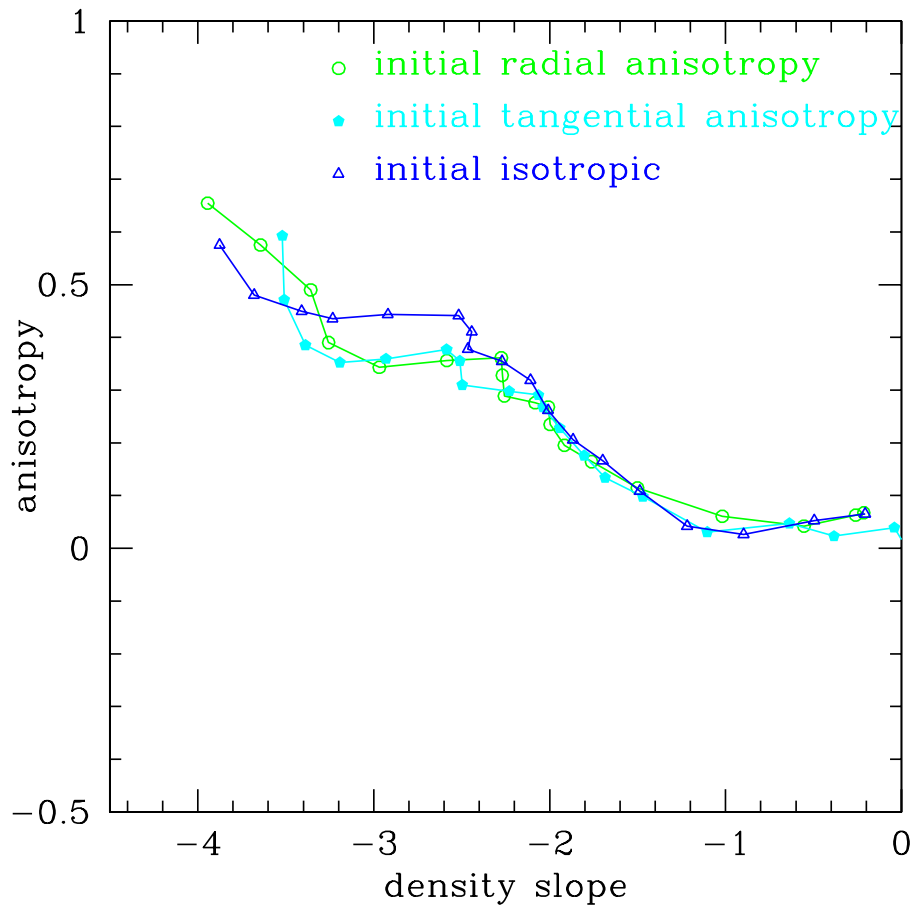}
\caption{Repeated head-on collision between two NFW structures, to
test the effect of different initial conditions. The different colours
refer to different initial conditions; green (open) circles show
the results from the initially radially anisotropic case (1), 
cyan (filled) pentagons show the results of the initially
tangentially anisotropic case (2), and blue (open) triangles was the 
isotropic collision described in section~\ref{sec:head}.  
In the region inside of a density slope of about $-2.4$, there is
virtually no different between the resulting $\ab$ relations.}
\label{fig:ic}
\end{center}
\end{figure}

We follow this with a simulation identical to the one just described, 
except that we place each individual particle in the initial structures on a
tangential orbit, without changing the energy of the individual particles. 
2a) This initial condition thus corresponds to strongly tangential motion. 
When letting this structure equilibrate in isolation it settles down
with a tangential anisotropy of $\beta \sim -2$.
2b) Again, two copies of this equilibrated structure are collided 
head-on as in step 1b.
The result of these tests are shown in figure~\ref{fig:ic}.

We see from figure~\ref{fig:ic} that the resulting $\ab$ relation is
virtually independent of initial conditions, and we conclude that
the collisions were sufficiently violent to erase the initial
conditions sufficiently to conform with the $\ab$ relation.
We emphasize, that we are not stating that initial conditions are
erased completely, however, that they are erased sufficiently to allow
the resulting structure to obey the $\ab$ relation.

\subsection{Symmetric simulations}

\begin{figure}[htb]
\begin{center}
\epsfxsize=13cm
\epsfysize=10cm
\includegraphics[width=0.8\textwidth]{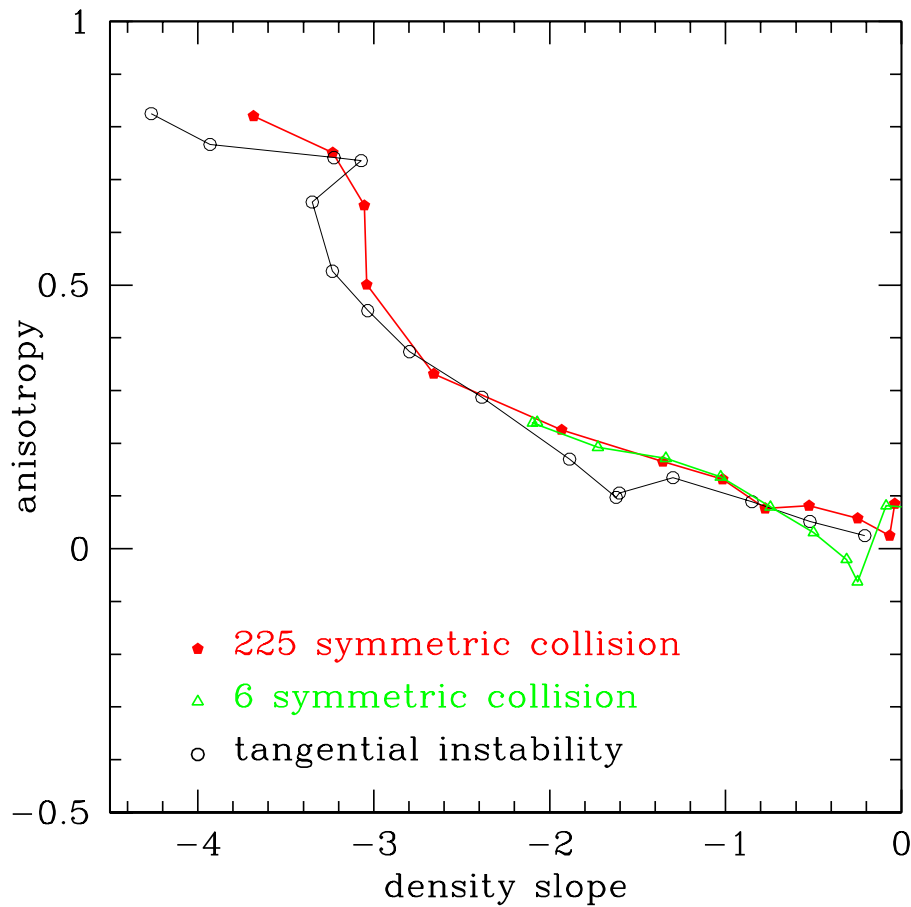}
\caption{Symmetric collision between 6 large isotropic NFW structures
(green open triangles); repeated collisions of numerous small
initially isotropic structures (red filled pentagons); tangential
instability (black open circles). These significantly different
simulations exhibit remarkably similar $\ab$ relations.}
\label{fig:sym}
\end{center}
\end{figure}

The simulations described above were all performed through very
non-symmetric collisions. We therefore performed two experiments
with symmetric collisions. First we take 6 copies of the
isotropic spherical NFW structure, each with 1 million particles,
and place them symmetrically along the x, y and z axes.  We collide
these with initial separation and relative velocities similar to 
what was described above. For this simulation we use softening 
of 2 kpc. The resulting structure has 6 million particles.
Second, we created an NFW structure with the same parameters
as described above, but only containing $10^4$ particles. We now
place 15 of these in a cubic symmetry (1 cell-centered, 6 face-centered
and 8 at corners) with initial separation about 600 kpc and initial 
relative velocities of 60 km/sec, and collide these together using 
softening of 1 kpc.  After letting
the resulting structure equilibrate we take this resulting structure,
place 15 copies in a cubic symmetry, and collide these again. The
resulting structure thus contains  2.25 million particles.
The results of these two simulations are shown in figure~\ref{fig:sym}.
There is good agreement between the resulting $\ab$ relation for these
two structures in the central region.

\subsection{Tangential instability}

One may fear that the resulting structures from the various simulations 
shown thus far end up near the same line in $\ab$ space simply because 
they sample only relatively similar kinds of merging processes.
The concern is that all the global properties of the final structures 
are the result of the fact that we slam structures into one another. 
Also the radial infall simulations included in ref.~\cite{hm} 
can be argued to be 
nothing but small structures falling into some larger structure.

In order to probe this issue further, we will simulate the evolution
of a rather artificial structure. We populate a small number of
particles, $4 \cdot 10^4$, uniformly (but with randomly chosen
positions) on an infinitely thin spherical shell of radius $0.01$ in
simulation units. These particles all have zero initial velocity.
Then we place 1 million particles on another infinitely thin spherical
shell at the radius $20$ in simulation units, again uniformly
distributed, but with randomly chosen positions.  These particles are
all placed on exactly circular orbits, but with random directions
(tangentially to this sphere). All particles have exactly the same
mass.  Now, if this system had infinitely many particles uniformly
distributed then the system would be in equilibrium, however, the
poissonian noise of the particles in the initial condition, and the
numerical noise from the integration of the equation of motion, imply
that the particles will slowly start clumping together at random
points on the sphere. The central collection of particles ($4 \cdot
10^4$),  which effectively act as a central massive mass point, 
quickly settles into a small equilibrated structure, leaving 
the gravitational influence on the outer particles virtually unaffected.
The particles on the outer shell, however, slowly break the
initial spherical symmetry. Particles passing nearby each other
start scattering off each other, kicking some particles outwards and
some particles inwards.  Eventually the entire system breaks up, and
the small initial central collection of particles are swallowed by
the large collection of particles from the outer shell.

We run this simulation with softening of 0.1 in these simulation
units, and we let the system evolve towards a new equilibrium state.
This state does, surprisingly enough, resemble an NFW structure in
density, and we extract the $\ab$ relation for this resulting
structure. The results are plotted as black (open) circles on
figure~\ref{fig:sym}, and the agreement with the other simulations is
striking.

We conclude that the appearance of the $\ab$ relation must have a 
deeper foundation than merely being the result of considering systems 
which were constructed to smash into each other.

\subsection{Comparison with previous works}

\begin{figure}[htb]
\begin{center}
\epsfxsize=13cm
\epsfysize=10cm
\includegraphics[width=0.8\textwidth]{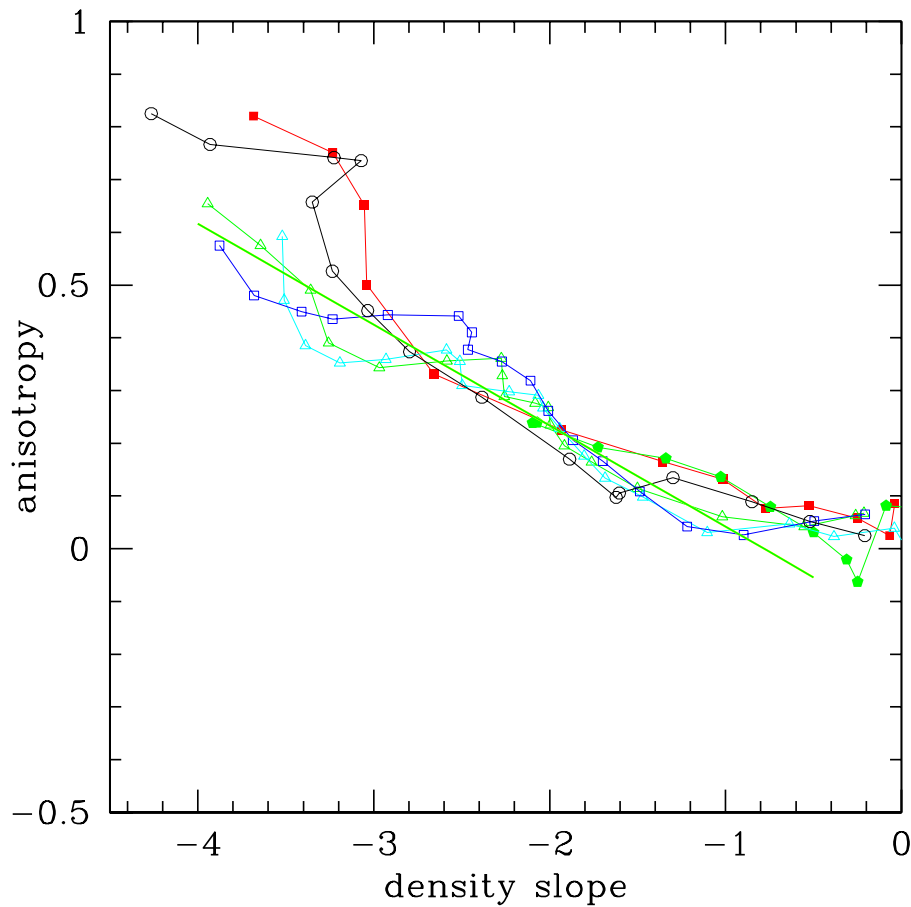}
\caption{A collection of the various simulations described in the
previous sections. The straight (green) line is from ref.~\cite{hm}, 
which was shown
to provide a reasonable fit to a collection of different simulations,
including mergers of disk galaxies, cosmological structures, and
radial infall simulations.}
\label{fig:together}
\end{center}
\end{figure}

In order to compare the results of these various simulations presented
above, we now plot the 3 double head-on collisions with different
initial conditions, the two symmetric collisions (between 6 and 225
initially isotropic NFW structures), and the tangential instability in
figure~\ref{fig:together}.  On the same figure we plot the line which
was argued to be a reasonable fit in ref.~\cite{hm}.  This line was a fit to
simulations including mergers of disk galaxies, cosmological
structures, and radial infall simulations, all simulations very
different from the ones considered in this paper.  We see indeed that
our simulations are in fair agreement with the results of ref.~\cite{hm}.

\section{Region of trust}
\label{sec:trust}

In all the figures above we have plotted a very extended region in the
density slope, which corresponds to plotting points very near the
central region (small negative $\alpha$) and very distant (large
negative $\alpha$).

\begin{figure}[htb]
\begin{center}
\epsfxsize=13cm
\epsfysize=10cm
\includegraphics[width=0.8\textwidth]{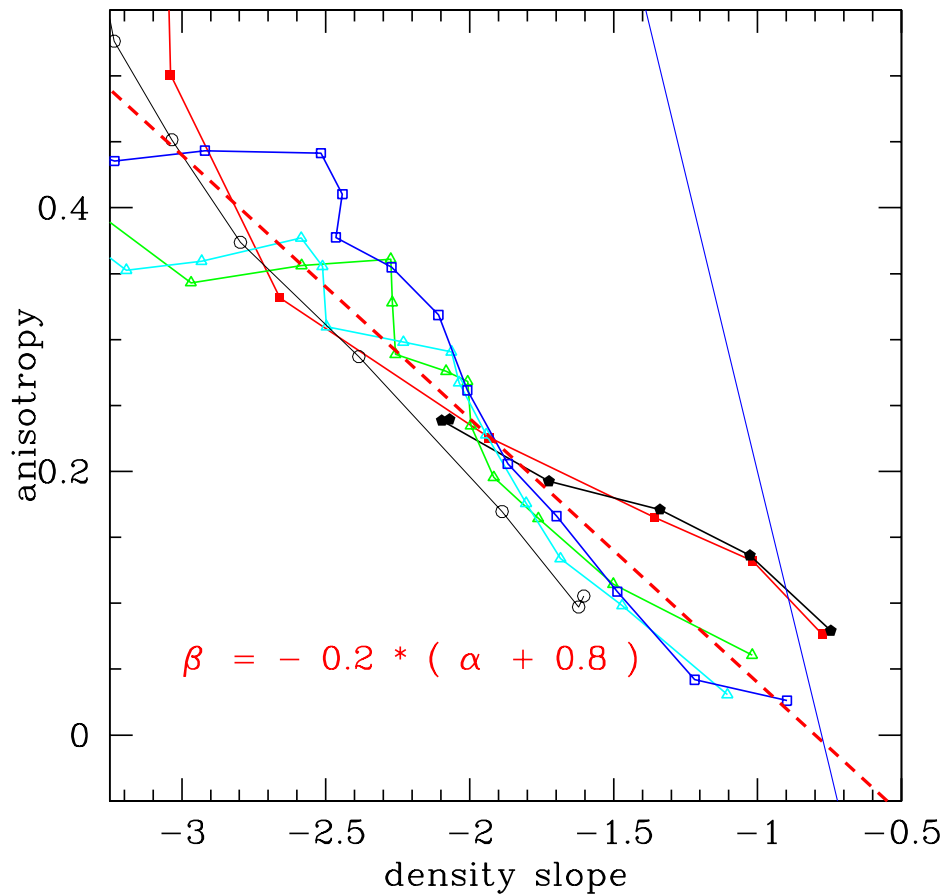}
\caption{Trusted region. We have excluded points inside a central 
region corresponding to 5 times the softening. We do not trust 
points further out than $\alpha = -3$. The red dashed line is a
fit to the points in this region, given by 
$\beta \approx -0.2 \, (\alpha+0.8)$.
The thin blue (solid) line shows the theoretical 
results of ref.~\cite{dm} for the connection between the central density
profile and the central anisotropy: from theory alone these
{\em central} values can be anywhere on this solid line.
The crossing of the two straight lines thus shows the actual
values for the central density slope and anisotropy.}
\label{fig:together2}
\end{center}
\end{figure}

In the central region one can generally only trust a region beyond
a few times the softening length. In almost all of our simulations 
this corresponds to a region where the density slope is about -1.
Further inwards numerical noise leads to density profiles more shallow 
than -1, which  cannot be trusted. However, most of our simulations
have been run for very long times (much longer than a 
corresponding typical cosmological simulation) and the numerical noise 
therefore has a slightly larger reach.
We are being rather conservative, and ignore the regions inside 5
times the softening. Concerning the outer region, we have 
visually inspected all the simulations, and it is clear that regions
with slopes more shallow than $-3$ are fully equilibrated. Looking
at figure~\ref{fig:together}, one also sees that the scatter in 
$\ab$ becomes large beyond $\alpha$ of $-3$. We therefore trust
the region with slopes more shallow than $ \alpha=-3$. 
This region (excluding points inside 5 times the softening length) is 
shown for each simulation in 
figure~\ref{fig:together2}, together with a line from the equation
\begin{equation}
\beta  = - 0.2 * ( \alpha  + 0.8 )
\label{eq:relation}
\end{equation}
which provides a reasonable fit within this region. We see that
the scatter in $\beta$ is only about $0.05$.

\section{Comparison with theory}

In a recent paper, Dehnen \& McLaughlin (2005) showed, that under
the two assumptions that phase-space density is a power-law 
in radius, and that there is a linear $\ab$ relation, that
the central slope of dark matter structures must be
$\alpha_0 = -(7 + 10\beta_0)/9$, where $\beta_0$ is the
central anisotropy.  In other words, the central values
of the density slope and the anisotropy must be somewhere on
the thin (blue, solid) line in figure~\ref{fig:together2},
however, from theory alone there is no telling where.

We can now compare this theoretical result with
our numerical findings. Our numerical results are approximated by the
fat (red, dashed) line.
We see that the two lines cross near $\beta=0$ and 
$\alpha=-7/9$, showing that the central part of dark matter structures
is isotropic, $\beta_0 \approx 0$,  
and that the central density slope is indeed $\alpha_0 \approx -7/9$.

Theory~\cite{austin} also 
shows that the outer density slope is about $\gamma_\infty
= -31/9 \approx -3.44$, which when compared with our findings result in 
$\beta_\infty \approx 0.53$.  Thus, from theoretical works we know
the innermost and outermost density slopes, and combined with the
results of this paper we now also know both innermost and outermost
anisotropy of pure dark matter structures.

Ref.~\cite{dm} noticed that when assuming that phase-space density is a
power-law in radius, then the Jeans equations have a particularly
simple form if and only if there is a {\em linear} $\ab$ relation as
originally suggested in ref.~\cite{hm}. 
Our numerical results indeed confirm
such linearity, and hence support the use of this particularly simple
version of the Jeans equation for theoretical studies.

\subsection{Comparison with cosmological simulations}

We clarified in section~\ref{sec:head} that the $\ab$ relation only
holds for systems which have been perturbed sufficiently and
subsequently allowed to relax. One can imagine setting up two
equilibrium systems each with zero anisotropy, and colliding these
with large impact parameter, in which case the system is not perturbed
sufficiently, and the resulting system should not be expected to land
near the $\ab$ relation.  As was shown in section~\ref{sec:head}, even
with zero impact parameter, only the densest part of the system
(inside density slope of about $-2.2$) will reach the $\ab$ relation
after only one collision. One should keep in mind that the problem
here is that setting up a system with zero anisotropy is very
artificial, and not in agreement with structures formed in
cosmological simulations.

Similarly, only the central part of structures formed in cosmological
simulations should be expected to land on the $\ab$ relation. This is
because most of the infalling matter has never been near the center of
the structure and hence has not been perturbed: most particles outside
half of the virial radius have not been through the center and have
therefore not been mixed sufficiently.  This expectation is indeed
confirmed by cosmological simulations (private communications with
J. Diemand and C. Power, see also the large scatter in figure 3 of
ref.~\cite{dm}) which show that the central regions of dark matter structure do
land on the $\ab$ relation, whereas the outermost regions do not quite.
The $\alpha-\beta$ relation has also been shown to hold for high-$\sigma$
subset of galaxy haloes~\cite{diemandmadau}.

\section{Attractor solution?}

It appears from the agreement between the different simulations
presented above, that the $\alpha-\beta$ relation is some kind of
{\em attractor}. In order to test such a claim one should perform minor
perturbations of the system, and then observe in which direction
the system flows. Antonov's laws of stability tell us that many
systems are in equilibrium, e.g. an isotropic Hernquist structure
will remain isotropic when exposed to minor perturbations. Antonov's
laws of stability are valid under the assumption that there is a zero
on the r.h.s. of the Boltzmann equation, and we will therefore 
expose our system to perturbations which act like an instantaneous
non-zero term on the r.h.s. of the Boltzmann equation.

\begin{figure}[htb]   
\begin{center}
\epsfxsize=13cm
\epsfysize=10cm
\includegraphics[width=0.8\textwidth]{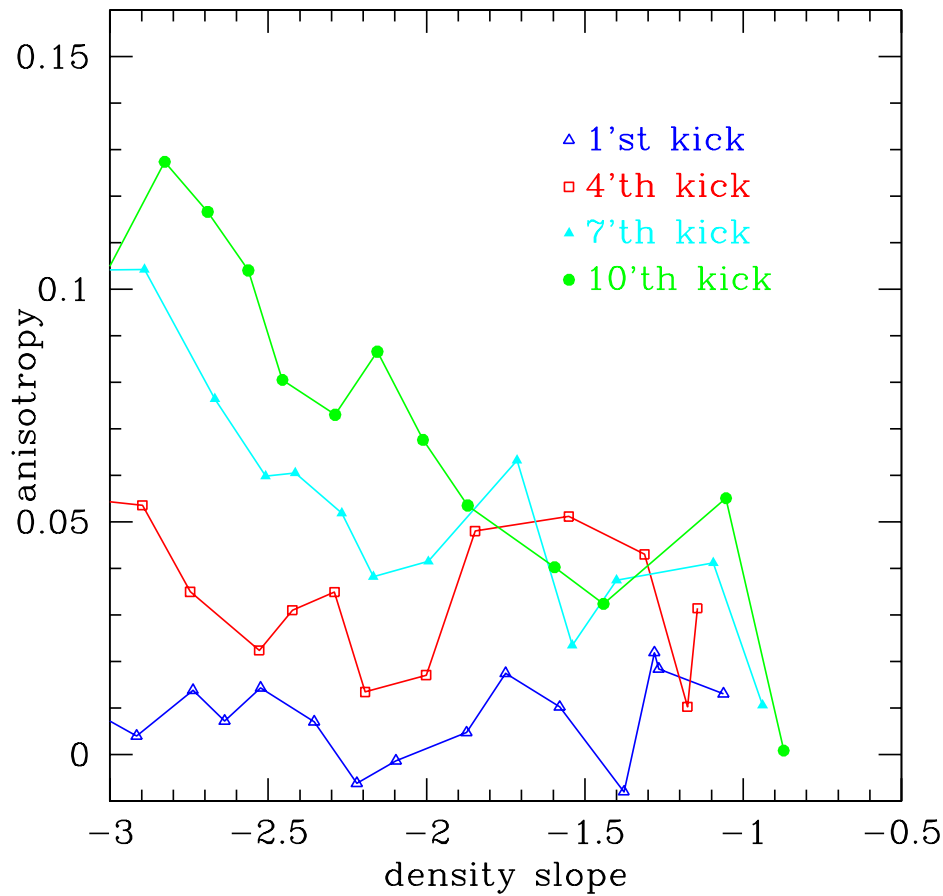}
\caption{We perturb the energies slightly of each individual particle
in an equilibrum system. After letting the system relax do we perturb
again. We show the resulting $\alpha-\beta$'s after 1, 4, 7 and 10
perturbations.  These perturbations are isotropic, and do therefore by
themselves not induce any anisotropy.  The system does indeed appear
to move towards the universal $\alpha-\beta$ relation, indicating that
it is indeed an attractor.}
\label{fig:kick}
\end{center}
\end{figure}

We take an isotropic NFW structure which is in total equilibrium.
Then we take each individual particle and change its velocity with a
random number, keeping its direction unchanged. This is done in such a
way that the new energy is changed by maximum $25\%$ from the original
energy, and in such a way that there is overall energy conservation.
This perturbation is thus completely isotropic, and should not by
itself change the isotropy of the system. Subsequently we let the
system relax to a new equilibrium state, and we extract density profile
and anisotropy of this new system.

We then take the new equilibrated system and perturb it in a similar
way with a new set of random numbers. After letting it relax, we can
repeat the process. The resulting $\alpha-\beta$'s are presented
in figure~\ref{fig:kick}, where we show the relaxed systems, after 
having perturbed 1, 4, 7 and 10 times. We see indeed, that the system
is slowly, but surely, moving in the direction of the universal
$\alpha-\beta$ relation. We therefore conclude that the 
$\alpha-\beta$ relation really is an attractor.

We did not perform further perturbations of the system, because for
anisotropic systems (like after the 10'th perturbation) these
isotropic perturbations will on their own tend to isotropize the
system, and hence the system may potentially appear to not flow
towards the attractor. Most probably one must devise more
sophisticated perturbations in order to see the flow all the way
toward the universal $\alpha-\beta$ line.

\section{Conclusions}

We have quantified the relationship between the density slope and the
velocity anisotropy, the $\ab$ relation. We have performed a large
set of simulations to investigate systematic effects related to 
shape and initial conditions, and we find that the $\ab$ relation is
almost blind to both shape and initial conditions, as long as the
system has been {\em perturbed sufficiently} and subsequently allowed
to relax.  We find strong indications that the relation is indeed
an attractor.

We have performed symmetric and highly non-symmetric simulations, 
along with a simulation of
the tangential orbit instability to extract the zero point and slope
of the $\ab$ relation in the regions which are numerically
trustworthy.  These simulations complement the simulations performed
in ref.~\cite{hm}, which included a cosmological simulation and radial infall
collapse, and yet exhibit a striking level of agreement with this work.
When compared with analytical results we find that the central region 
is indeed isotropic, and that the outer asymptotic anisotropy is radial, 
with a magnitude of $\beta \approx 0.5$.

\ack 
It is a pleasure to thank Ben Moore for discussions and encouragement.
SHH thanks the Tomalla foundation for support.

\label{lastpage}


\section*{References}
\begin{thebibliography}{99}

\bibitem{nfw96}
Navarro J F, Frenk C S and  White, S D M,   
1996 ApJ, 462, 563

\bibitem{moore}
Moore B et al.,
1998 ApJ, 499, 5


\bibitem{Fukushige1997}
Fukushige T and Makino J,
1997 ApJ, 477, L9 


\bibitem{Moore1998}
Moore B et al.,
1998
ApJ, 499, L5

\bibitem{Moore1999pro}
Moore B et al.,
1999
MNRAS, 310, 1147 

\bibitem{Ghigna1998}
Ghigna S et al.,
1998
MNRAS,  300, 146



\bibitem{Fukushige2004}
Fukushige T, Kawai A and Makino J, 
2004 ApJ, 606, 625




\bibitem{Tasitsiomi2004}
Tasitsiomi A et al.,
2004
ApJ, 607, 125 


\bibitem{Navarro2004}
Navarro J et al.,
2004 MNRAS, 349, 1039

\bibitem{Reed2004}
Reed D et al.,
2005 MNRAS, 357, 82


\bibitem{diemand04}
{Diemand} J, {Moore} B and  {Stadel} J,  
2004  353, 624

\bibitem{power}
Power C et al.,
2003 MNRAS, 338, 14


\bibitem{Diemand:2005wv}
Diemand J et al.,
2005 astro-ph/0504215.



\bibitem{cole} 
Cole S \& Lacey C,
1996
MRNAS, 281, 716

\bibitem{carlberg} 
Carlberg R G,
1997 ApJ, 485, L13



\bibitem{taylor}
Taylor J E and Navarro J F,
2001 ApJ, 563, 483


\bibitem{hm}
Hansen S H and Moore B, 
2005 New Astron.\ (to appear) 
astro-ph/0411473


\bibitem{students}
Hansen S H, Egli D, Hollenstein L and Salzmann C,
2005 New Astron.\   10, 379

\bibitem{HMZS}
Hansen S~H, Moore B, Zemp M and Stadel J,
2005  arXiv:astro-ph/0505420



\bibitem{jeanspaper}
Hansen S~H, 
2004 MNRAS, 352, L41


\bibitem{austin}
Austin C~G et al.,
2005 astro-ph/0506571.


\bibitem{dm}
Dehnen W and McLaughlin D,
2005 astro-ph/0506528



\bibitem{ascasibar}
Ascasibar Y, Yepes G, Gottl\"ober S and M\"uller V,
2004 MNRAS, 352, 1109


\bibitem{rasia}
Rasia E, Tormen G and Moscardini L,
2004 MNRAS, 351, 237

\bibitem{barnes}
Barnes  E~I et al.,
2005 astro-ph/0510332.


\bibitem{stelios}
Kazantzidis S, Magorrian J and Moore B, 
2004 ApJ, 601, 37


\bibitem{stadel01}
Stadel J,
2001
PhD thesis, University of Washington



\bibitem{parisconf}
Hansen S H, Moore B and Stadel J,
2005
Proceedings of 21st IAP Colloquium, Paris


\bibitem{merritt85}
Merritt D,
1985 MNRAS, 217, 787


\bibitem{diemandmadau}
Diemand J, Madau P and Moore B,
2005 astro-ph/0506615.



\end{thebibliography}
\end{document}